\documentclass[pre,twocolumn,floatfix,bibnotes]{revtex4}
\usepackage{here}
\usepackage[dvips]{graphicx}

\usepackage{bm}

\begin{document}

\title{Tunable transmission and bistability in left-handed bandgap
  structures}

\author{Michael W. Feise}
\affiliation{Nonlinear Physics Centre, Centre for Ultra-high
bandwidth Devices for Optical Systems (CUDOS), Research School of
Physical Sciences and Engineering, Australian National University,
Canberra, ACT 0200, Australia}

\author{Ilya V. Shadrivov}
\affiliation{Nonlinear Physics Centre,
 Centre for Ultra-high
bandwidth Devices for Optical Systems (CUDOS), Research School of
Physical Sciences and Engineering, Australian National University,
Canberra, ACT 0200, Australia}

\author{Yuri S. Kivshar}
\affiliation{Nonlinear Physics Centre, Centre for Ultra-high
bandwidth Devices for Optical Systems (CUDOS), Research School of
Physical Sciences and Engineering, Australian National University,
Canberra, ACT 0200, Australia}

\date{\today}

\begin{abstract}
We study the defect-induced nonlinear transmission of a periodic
structure created by alternating slabs of two materials with
positive and negative refractive index. We demonstrate bistable
switching and tunable nonlinear transmission in a novel type of
bandgap that corresponds to the vanishing average refractive
index, and compare the observed effects for two types of the
bandgaps.
\end{abstract}

\maketitle

Materials with both negative electric permittivity and negative
magnetic permeability were suggested theoretically a long time
ago~\cite{Veselago:SPU-10-509} and they were termed {\em
left-handed materials}~\cite{Pendry:2003-639:OE}.  Such materials
can also be described by a negative refractive index, as was
demonstrated by several reliable
experiments~\cite{Shelby:SCIENCE-292-1-77,Greegor:2003-688:OE} and
numerical finite-difference time-domain
simulations~\cite{Foteinopoulou:2003-107402:PRL}.

Multilayered structures composed of left-handed (LH) materials can
be considered as a sequence of flat lenses that provide an optical
cancellation of the conventional right-handed (RH) layers leading
to either enhanced or suppressed
transmission~\cite{Zhang:2001-1097:APL,Pendry:2003-6345:JPCM}.
More importantly, a one-dimensional stack of layers with
alternating RH and LH material with vanishing average
refractive index $\mathopen{<}n\mathclose{>}$ displays a novel
bandgap~\cite{Li:PRL-90-083901,Shadrivov:2003-3820:APL,Wu:2003-235103:PRB}
that differs from a conventional Bragg reflection gap.

\begin{figure}[tbp]
  \centerline{\includegraphics[width=3.3in]{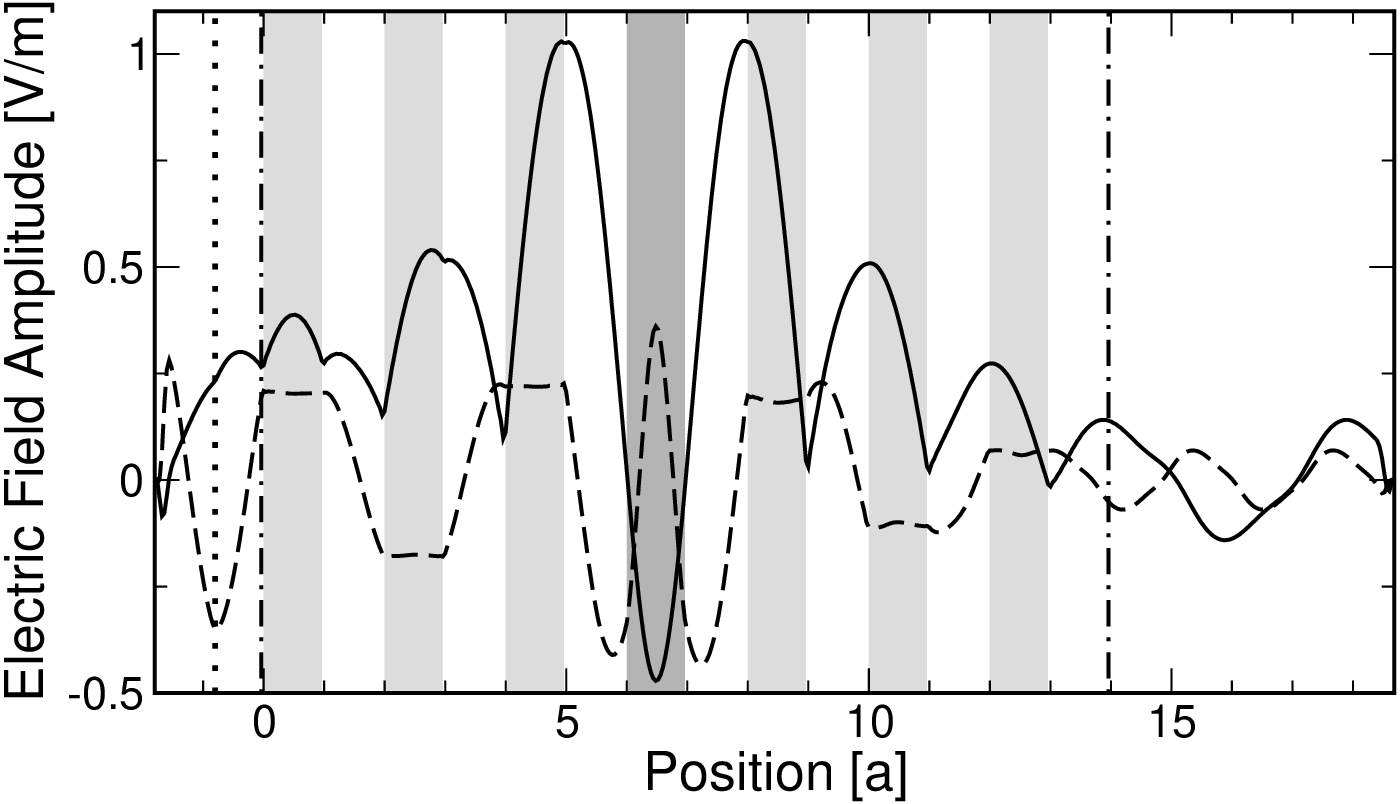}}
  \caption{
  Layout of the system with snapshots of the electric field
    amplitude. A Gaussian pulse with carrier frequencies
    $\omega_c=2\pi\times 14.93$~GHz (solid) and $\omega_c=2\pi\times
    25.88$~GHz (dashed) is incident on the structure. The fields are
    shown at the pulse peak. The shaded background indicates the
    material structure, with dark grey denoting the Kerr material and
    light grey the LH slabs. The defect layer is centered within the
    structure.  The dotted vertical line shows the source location,
    the dash-dotted vertical lines denote the
    boundaries of the considered structure. The field is incident from
    the left.
    }
  \label{fig:field-snapshots}
\end{figure}

In this letter, we study defect-induced bistable switching in the
novel bandgap structures and, for the first time to our knowledge,
compare the effects observed for zero-index and Bragg reflection
gaps. The particular structure we study by numerical
pseudospectral time-domain (PSTD) simulations consists of seven
periods of a LH-RH double-layer (indicated in
Fig.~\ref{fig:field-snapshots}). Each individual layer has equal
width $a$.  The LH layer of the central period is replaced by a
Kerr-type nonlinear material, which constitutes a structural
defect of the periodic system even in the linear regime. We also
study the effects produced by a change of the defect position for
two types of bandgaps.

We calculate the electric and magnetic fields directly from
Maxwell's equations using the PSTD method~\cite{Liu:MOTL-15-158}.
The differential Maxwell equations are approximated by difference
equations and solved iteratively. The material properties are
treated through the electric permittivity $\varepsilon_r$ and
magnetic permeability $\mu_r$. The PSTD method is advantageous for
the modeling of interfaces where both $\varepsilon_r$ and $\mu_r$
change because it samples all material properties at the same
location~\cite{Feise:TAP-submitted-2003}.

The linear properties of the LH material are described by
Lorentz dispersion characteristics in both $\varepsilon_r$ and
$\mu_r$,
\begin{eqnarray}
  \label{eq:1}
  \varepsilon_r(\omega) &=&
  1+\frac{\omega_{pe}^2}{\omega_{1e}^2-\omega^2-i\gamma_e\omega},\\
  \label{eq:2}
  \mu_r(\omega) &=&
  1+\frac{\omega_{pm}^2}{\omega_{1m}^2-\omega^2-i\gamma_m\omega}.
\end{eqnarray}
These functions are substituted into the relations,
$\mathbf{D}(\omega)=\varepsilon_r(\omega)\varepsilon_0\mathbf{E}(\omega)$,
$\mathbf{B}(\omega)=\mu_r(\omega)\mu_0\mathbf{H}(\omega)$, transformed
into ordinary differential equations in the time domain, and
subsequently approximated by difference
equations~\cite{Hulse:JOSAA-11-1802} that are incorporated into the
PSTD algorithm. This allows one to model frequency dependent
material properties in PSTD simulations. The nonlinear Kerr
material is described by a dielectric function
\begin{equation}
\label{eq:4}
  \varepsilon_r^{\mathrm{NL}}=1+\chi^{(1)} +
  \chi^{(3)}\left|\mathbf{E}(t)\right|^2.
\end{equation}
There are several established methods of including the nonlinear
material response in the PSTD
algorithm~\cite{Goorjian:OL-17-180,Ziolkowski:TAP-45-375,Lixue:OC-209-491,Tran:OL-21-1138}.
Here, we use the electric field of the previous time step to
evaluate
Eq.~(\ref{eq:4})~\cite{Ziolkowski:TAP-45-375,Lixue:OC-209-491}.

We also calculate the properties of the structure using the
transfer-matrix method (TMM)~\cite{Yeh:OWLM-1988}.  This method allows an
exact analytical solution of the linear problem. In particular, one
can relate incident, transmitted and reflected fields using the
transfer matrix of the structure, and thus obtain an explicit expression
for the transmission and reflection coefficients.

We describe the LH material by Eqs.~(\ref{eq:1}),(\ref{eq:2}) with
parameters chosen to give a refractive index $n\approx-1$ at the
design frequency $f_0=15$~GHz.  We use
$\omega_{pe}=1.1543\times 10^{11}$~s$^{-1}$,
$\omega_{1e}=\omega_{1m}=2\pi\times 5\times 10^6$~s$^{-1}$, and
$\omega_{pm}=1.6324\times 10^{11}$~s$^{-1}$, and include small
losses, $\gamma_e=2\times\gamma_m=2\pi\times 6\times
10^6$~s$^{-1}$.  With these parameters the LH meta-material is
left-handed for frequencies $f<18.5$ GHz and right-handed for
$f>26$ GHz.
                                %
The slab thickness $a$ is chosen to be $\lambda_0/4$, where
$\lambda_0$ is the free-space wavelength at $f_0$. We use air as
the RH medium.  The parameters for the Kerr material are
$\chi^{(1)}=3$ and $\chi^{(3)}=4$ $\mathrm{m^2/V^2}$.  In each case the PSTD
simulations use a discretization of 100 points per $\lambda_0$ and a time step corresponding to half the
Courant stability limit~\cite{Liu:MOTL-15-158} of the linear case.

\begin{figure}[tbp]
  \centerline{\includegraphics[width=3.3in]{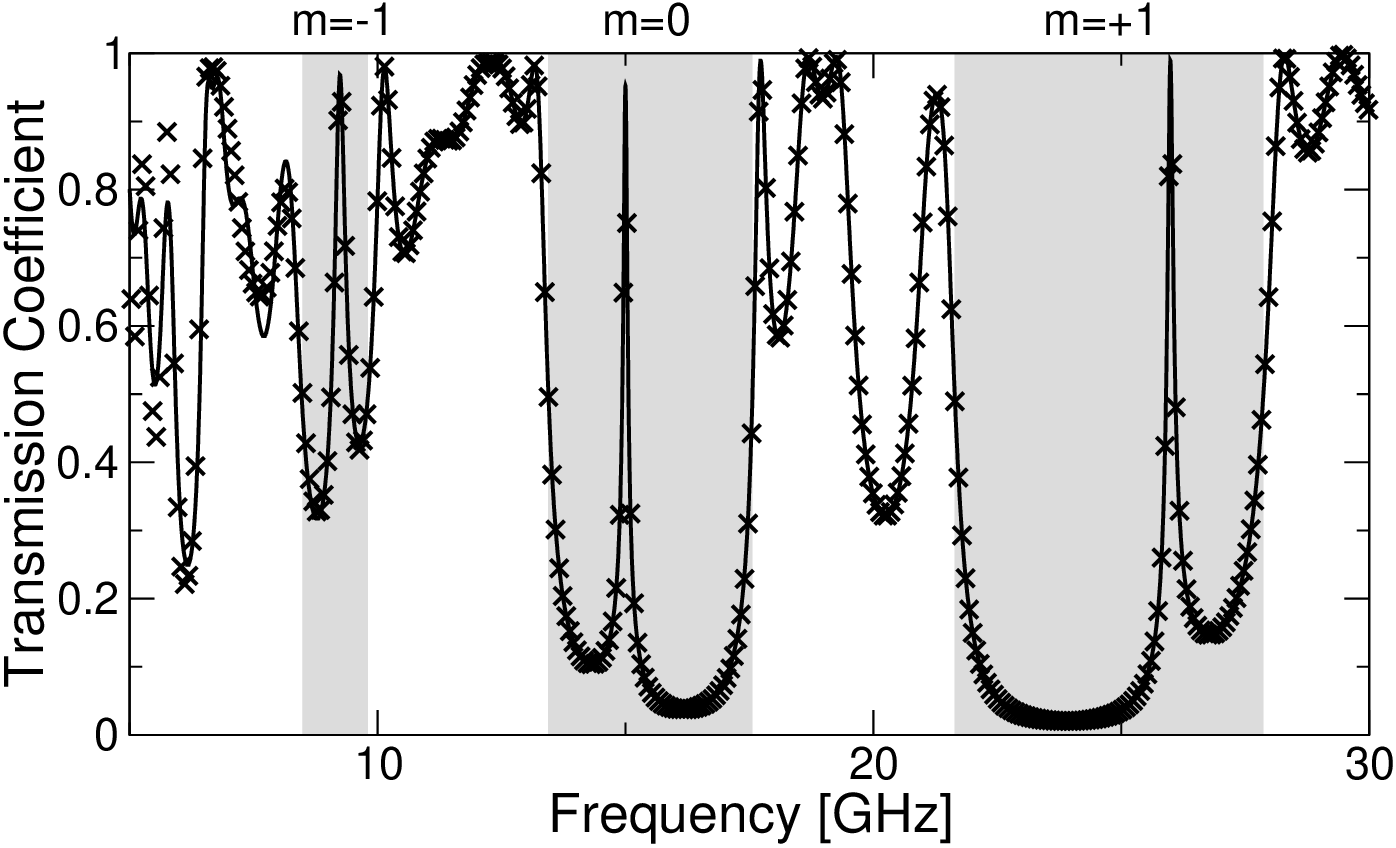}}
  \caption{Linear amplitude transmission coefficient of the
    structure calculated by the TMM (line) and PSTD (symbols) methods.
    The band gaps are indicated by shaded regions.  The
    $\mathopen{<}n\mathclose{>}=0$ ($m=0$)
    bandgap lies around 15 GHz.
    The first conventional Bragg gap ($m=+1$)lies around 25 GHz while the
    first Bragg gap in the left-handed regime ($m=-1$) lies around 9 GHz.
  }
  \label{fig:linear_transmission_spectrum-1}
\end{figure}

To validate our PSTD calculations, we first study the amplitude
transmission spectrum in the linear regime, shown in
Fig.~\ref{fig:linear_transmission_spectrum-1}, and find excellent
agreement between the TMM and PSTD methods. For frequencies below
8~GHz, the discretization level in the PSTD simulations is
insufficient for accurate results, due to the high refractive
index of the LH material, and some discrepancies appear. The
structure exhibits band gaps due to Bragg scattering, both in the
RH and in the LH frequency region. An additional band gap appears
around the frequency where the average refractive index vanishes,
as was addressed
earlier~\cite{Li:PRL-90-083901,Shadrivov:2003-3820:APL,Wu:2003-235103:PRB}.
These band gaps can be identified through the index $m$ in the usual
Bragg condition,
\begin{equation}
  \label{eq:5}
  k_{\mathrm{RH}}a_{\mathrm{RH}} + k_{\mathrm{LH}}a_{\mathrm{LH}}
  = m\pi,
\end{equation}
where $k_j$ is the respective wave number and $a_j$ is the
respective layer thickness. The band gap index $m$ can be any
integer, including zero and negative numbers. We find that the
defect layer introduces a transmission peak into {\em each of the
shown band gaps}. The remainder of this letter will focus on the
comparison of the $\mathopen{<}n\mathclose{>}=0$ gap ($m=0$) and
the first RH Bragg gap ($m=+1$).

\begin{figure}[tbp]
  \centerline{\includegraphics[width=3.3in]{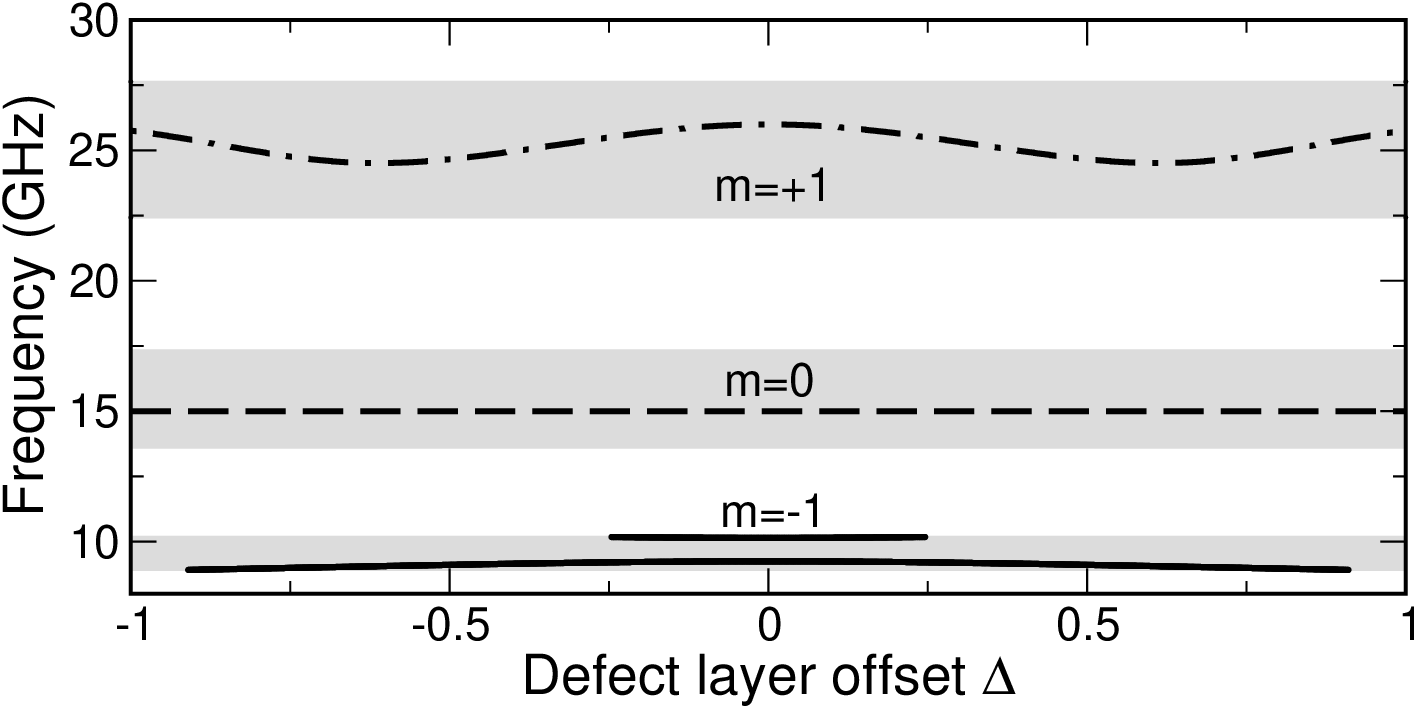}}
  \caption{
    Dependence of the defect frequency on the position offset $\Delta$
    of the defect layer in the linear limit.}
  \label{fig:freq-vs-position}
\end{figure}

We calculate the frequency of the defect modes in the band gaps
for different locations of the defect layer in our structure in
the linear case, as shown in Fig.~\ref{fig:freq-vs-position}. The
defect layer is shifted by a quantity $a\Delta$ ($|\Delta|\leq 1$) from the central
position ($\Delta=0$) between its two neighboring LH layers
without changing the rest of the structure.  It is remarkable that
in the $m=0$ gap the defect frequency {\em does not depend on the
defect position} while the one in the $m=+1$ gap is shifted with
changing $\Delta$.   When the thickness of the defect layer is not
equal to $a$ (not shown), the $m=0$ defect frequency also depends
on $\Delta$. This dependence, however, is much weaker than that
of the $m=+1$ defect.

In the nonlinear regime, the behavior of the fields depends on the
intensity of the electric field inside the nonlinear layer. We
study this problem by PSTD simulations.  The incident
field has a Gaussian envelope in time with a width parameter of
$1528/f_0$ and amplitude 0.2~V/m. The carrier frequency is
$\omega_c=2\pi\times 14.93$~GHz for the
$m=0$  band gap and $\omega_c=2\pi\times
25.88$~GHz for the $m=+1$ gap.  These frequencies lie below
their respective defect frequencies at $\Delta=0$ with equal
relative detuning.

\begin{figure}[tbp]
  \centerline{\includegraphics[width=3.3in]{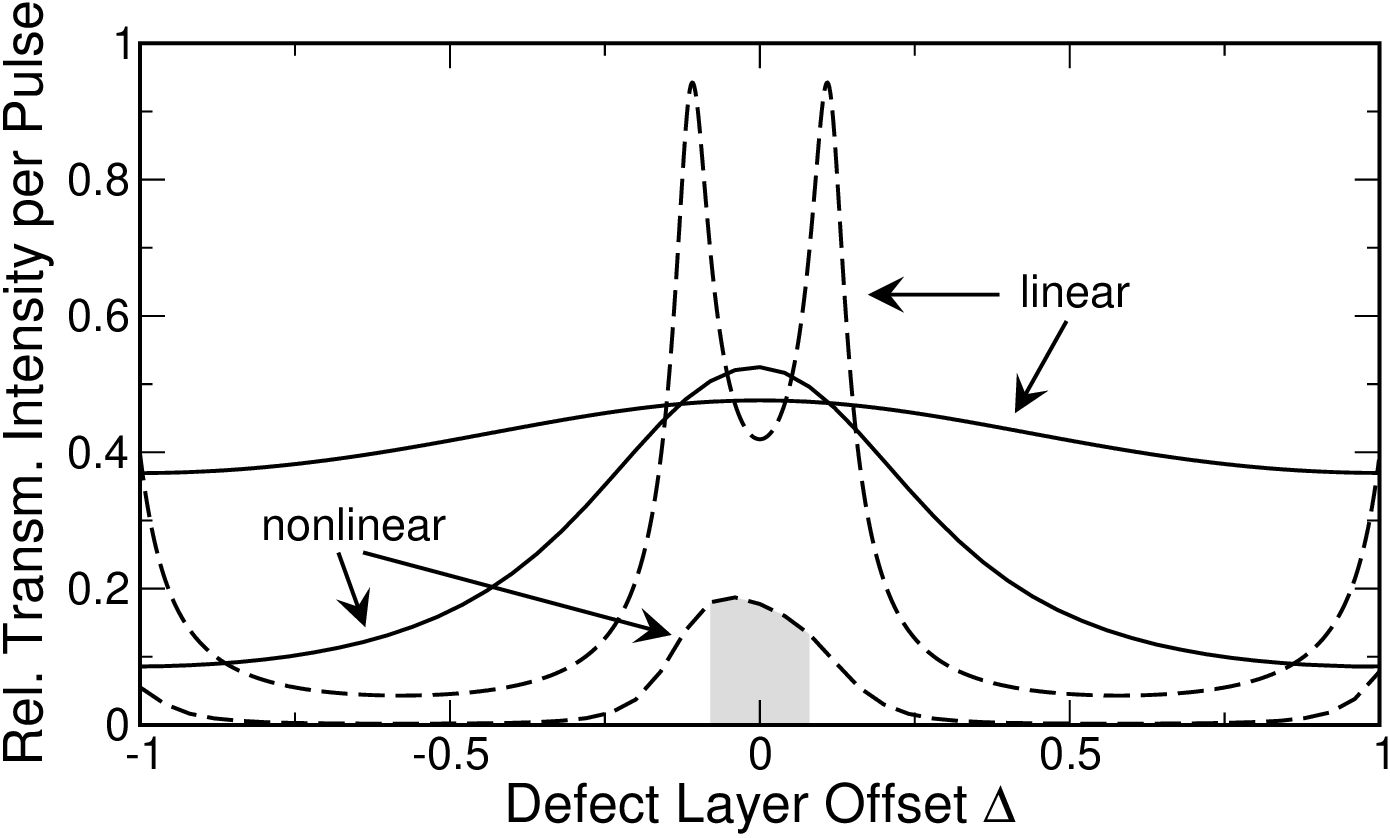}}
  \caption{
    Time-integrated transmitted intensity per pulse, relative to that
    of the incident pulse, in the $m=0$ gap (solid) and the $m=+1$ gap
    (dashed). Bistability occurs in the $m=0$ gap for all offsets,
    while the bistability region for the $m=+1$ gap is indicated in
    the graph by grey shading.
  }
  \label{fig:transmission}
\end{figure}
In Fig.~\ref{fig:transmission} we show the relative
time-integrated transmitted intensity per pulse versus the
position of the nonlinear layer in the {\em linear} and {\em
nonlinear} regimes for pulses at both carrier frequencies. The two
bands display very different character.  In the linear regime, the
transmission in the $m=0$ gap is only weakly dependent on $\Delta$
with a maximum at $\Delta=0$.  In the $m=+1$ gap, the linear
transmission shows a double peak structure near $\Delta=0$. Away
from this region the transmission is much reduced.  For
$\Delta\approx \pm 1$ the transmission increases again. In the
nonlinear case the transmission is generally reduced. In the $m=0$
gap, it maintains its character with a single peak in the center,
which actually exceeds the peak of the linear case. In the $m=+1$
gap, the double peak character is lost and a much smaller single
peak in the center and weak wings for large layer offsets remain.
For all other $\Delta$ the transmission almost vanishes.

\begin{figure}[tbp]
  \centerline{\includegraphics[width=3.3in]{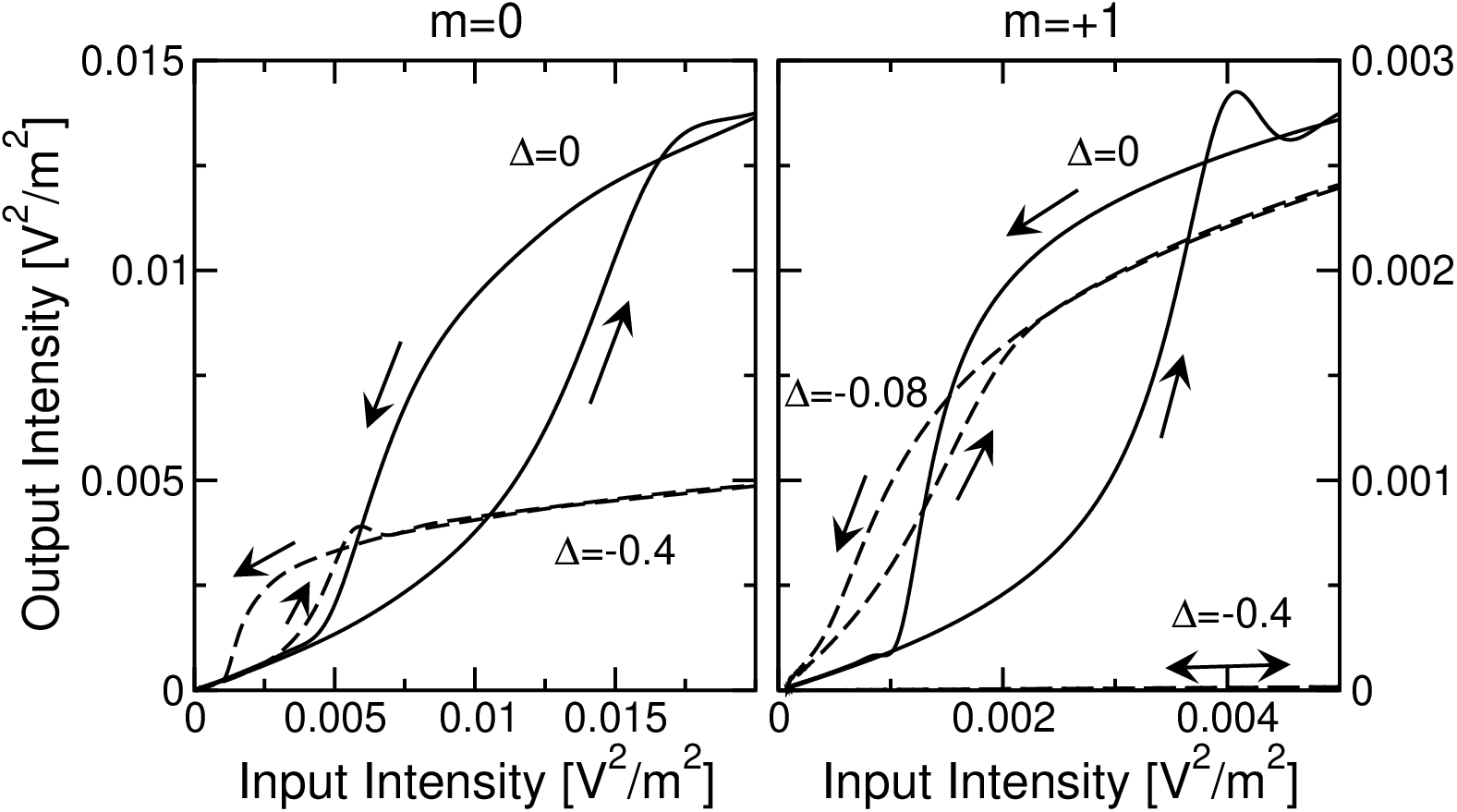}}
  \caption{
Output vs. input intensity of the structure, showing hysteresis
behavior for defect offsets of $\Delta=0, -0.4$ in the
$m=0$   gap and $\Delta=0, -0.08, -0.4$ in
the $m=+1$ gap. The $\Delta=-0.4$ curve in the $m=+1$ gap is an
almost vanishing straight line without bistability.
  }
  \label{fig:bistability}
\end{figure}

One of the most promising applications of periodic structures with
embedded nonlinear defect layers is the possibility to achieve
intensity-dependent tunable transmission in the spectral bandgaps
usually associated with bistability. Optical bistability is a
powerful concept that could be explored to implement all-optical
logic devices and switches. In nonlinear systems that display
bistability, the output intensity is a strong nonlinear function
of the input
intensity~\cite{Mingaleev:2002-2241:JOSB,Yanik:2003-2739:APL}.

In the $m=0$ gap, we observe bistability for all offsets while in
the $m=+1$ gap, bistability is observed for very small values of
$|\Delta|$. These effects are related to the behavior of the
defect frequency in the $m=+1$ gap with different $\Delta$. In this
gap the defect frequency shifts away from the carrier
frequency with increased $\Delta$ and eventually approaches it
again. In contrast to this, the defect frequency in the $m=0$ gap
stays very near to the carrier frequency for all layer offsets. In
Fig.~\ref{fig:bistability} we show the output vs.\ input intensity
during the time of the pulse. The left panel shows the behavior in
the $m=0$ gap for offsets $\Delta=0$ and $\Delta=-0.4$, while the
right panel shows the behavior in the $m=+1$ gap for offsets
$\Delta=0$, $\Delta=-0.08$, and $\Delta=-0.4$.  The bistable
behavior, where the output intensity with increasing input
intensity is different from the one with decreasing input
intensity, is clearly visible. The threshold switching intensity
in the $m=+1$ gap is significantly lower than in the $m=0$ gap. On
the other hand, the difference between switch-up and switch-down
threshold is much larger in the $m=0$ gap. The difference between
the behavior in the two gaps is illustrated by the $\Delta=-0.4$
curves.  While the $m=0$ gap shows significant bistability, the
$m=+1$ gap shows no bistability and almost no transmission.

In conclusion, we have studied the defect-induced nonlinear
transmission and bistability for two types of bandgaps of a
layered structure composed of two materials with positive and
negative refractive index with a Kerr nonlinear defect. We
demonstrated a number of unique features of the bandgap associated
with zero average refractive index of the structure.


\end{document}